\documentclass[aps,pre,reprint,twocolumn,showpacs,amsmath,amssymb]{revtex4-1}

\usepackage{graphicx}
\usepackage{bm}
\usepackage{color}
\usepackage{lineno}

\begin{document}

\title{Steady streaming viscometry of Newtonian liquids in microfluidic devices}

\author{Giridar Vishwanathan}
\author{Gabriel Juarez}
\email{gjuarez@illinois.edu} 

\affiliation{Department of Mechanical Science and Engineering, University of 
Illinois at Urbana-Champaign, Urbana, Illinois 61801, USA}

\date{\today}


\begin{abstract}

We report a novel technique capable of measuring the kinematic shear viscosity of 
Newtonian liquids with steady streaming in microfluidic devices. This probe-free 
microrheological method utilizes sub-kilohertz liquid oscillation frequencies around 
a cylindrical obstacle, ensuring that the inner streaming layer is comparable in size 
to the cylinder radius. To calibrate the viscometer, the evolution of the inner streaming 
layer as a function of oscillation frequency for a liquid of known viscosity is 
characterized  using standard particle tracking techniques. Once calibrated, we show 
how the steady streaming viscometer can be used to measure low-viscosity liquids and 
volatile liquids.

\end{abstract}

\maketitle


Steady streaming flows have received a renewed interest over the past decade due 
to their numerous applications in microfluidic devices \cite{Friend, Wiklund}.
Here, steady streaming refers to the rectified flow \cite{Riley} that occurs near 
the boundary of a rigid body of length-scale $a$, oscillating with frequency $f$ 
and small-amplitude $s$ ($\ll a$), in a stationary incompressible fluid of kinematic 
viscosity $\nu$. The magnitude of the characteristic streaming velocity 
scales as $U_s \sim \epsilon s \omega$, where $\epsilon = s/a$ is the dimensionless 
amplitude and $\omega = 2 \pi f$ is the angular frequency. At small scales, these 
rectified flows have been shown to be  useful in non-contact manipulation \cite{Amit}, 
trapping \cite{Marmottant,Lutz02,Lieu,Yazdi}, and sorting \cite{Wang,Thameem} of 
particles and cells as well as in enhancing pumping \cite{Girardo} and mixing 
\cite{Sritharan,Lutz,Ahmed} at low Reynolds numbers. There are opportunities to 
use steady streaming for the rheology of liquids as well \cite{Vlassopoulos}.

Microrheology aims at measuring local material properties of small quantities of
fluids by studying the relationship between deformation and stress. Optical 
microrheology techniques rely on the tracking of flow tracers under passive thermal 
fluctuations or active external forcing \cite{Waigh,Squires,Wilson,Zia}. The development 
of microrheological methods for complex fluids and soft materials \citep{Pipe,Waigh16} 
has been motivated by the advantages offered by microfluidic devices compared to 
conventional bulk techniques, such as small sample volume, reduction of free-surface 
effects, direct visualization of the underlying microstructure, and the ability to 
quantify low-viscosity and weakly viscoelastic solutions 
\cite{DelGiudice2,DelGiudice3}. These specialized methods have been shown to 
measure material properties such as the steady shear viscosity \cite{Gupta}, the 
most widely characterized material property, as well as the extensional viscosity 
\cite{Galindo, Haward}, and the longest relaxation time \cite{DelGiudice,Zilz} of 
various fluids. 

\begin{figure}[b]
\centering
	\includegraphics[width=\linewidth]{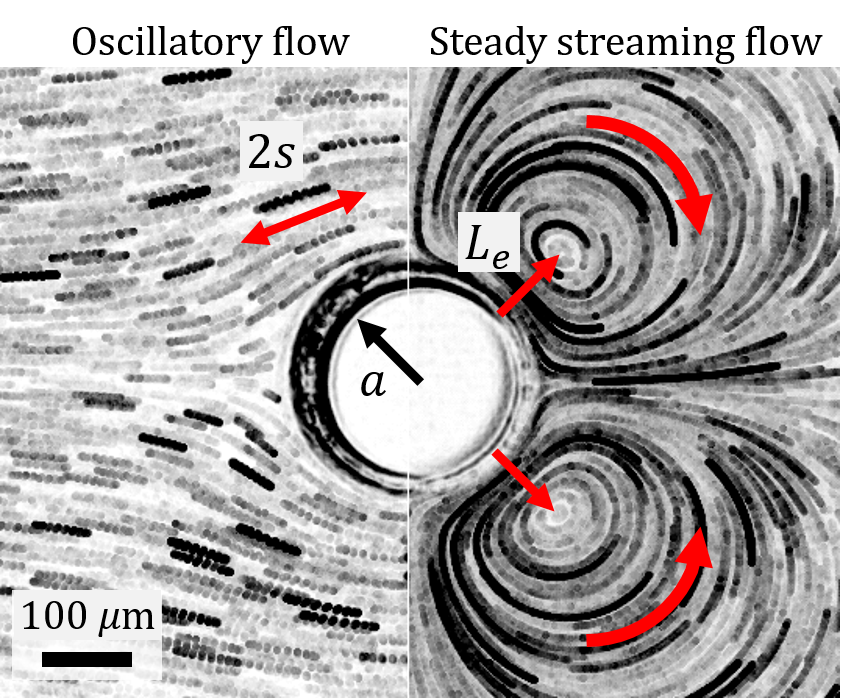}
\caption{Steady streaming in microfluidic devices. (Left) Pathlines of tracer 
particles captured with high-speed imaging. The displacement amplitude of $2s$ 
is shown as the fluid undergoes one period of oscillation. (Right) Pathlines of 
tracer particles captured with stroboscopic imaging. Half of the steady streaming 
profile is shown with two counter-rotating vortices and the location of the eddy 
center, $L_e$. Here, the cylinder radius is $a = 100 \ \mu$m. See supplementary 
movies.}
	\label{fig:figone}
\end{figure}

\begin{figure*}
\centering
	\includegraphics[width=\linewidth]{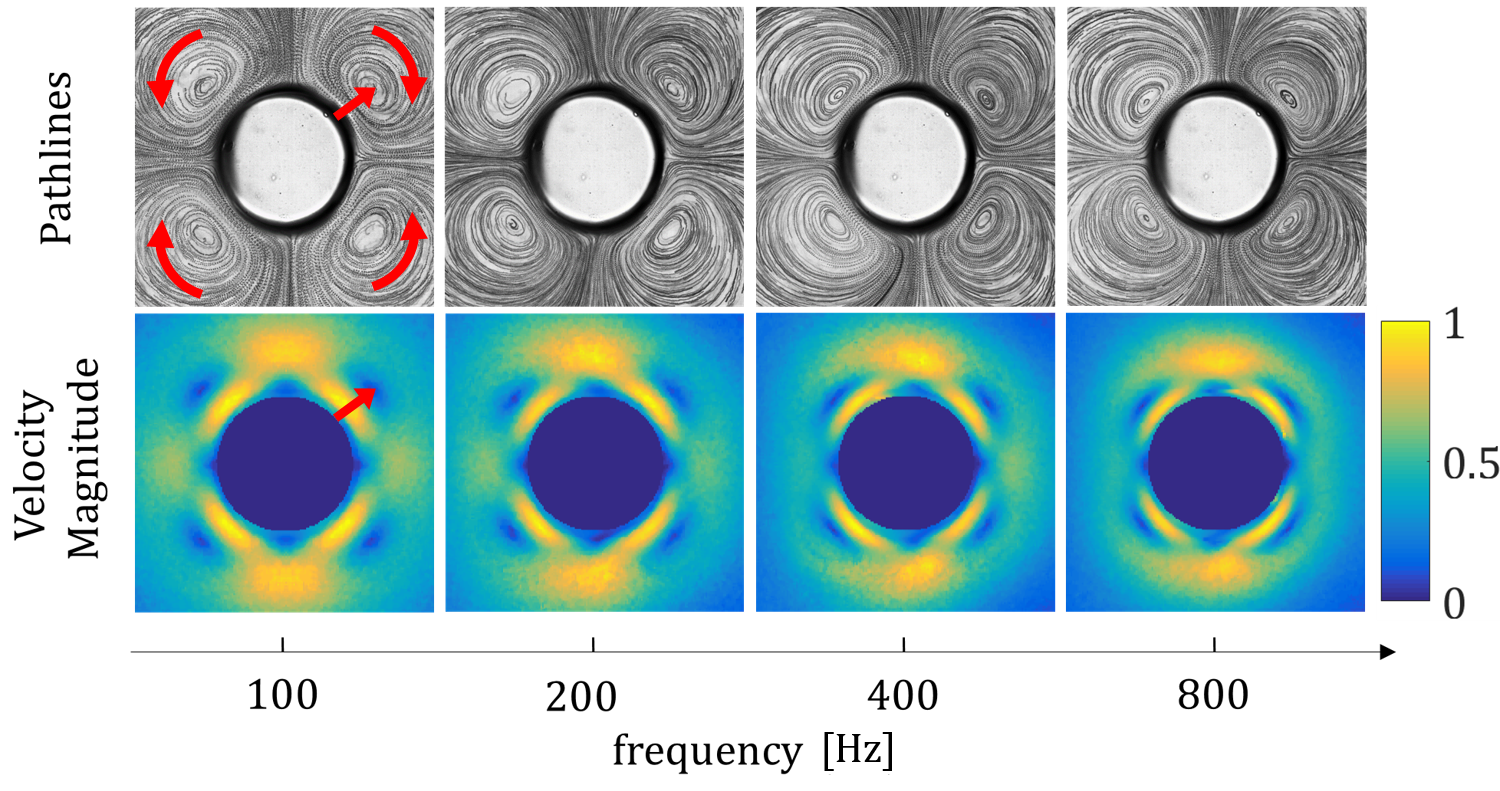}
\caption{Evolution of the steady streaming profile with increasing oscillatory 
frequency. (top) Pathlines of tracer particles captured with stroboscopic imaging
showing the steady streaming profile around a cylindrical obstacle. The location of 
the eddy center moves closer to the cylinder boundary as the oscillation frequency 
is increased. (bottom) Normalized velocity field magnitude profiles obtained 
from particle tracking velocimetry (color online).}
	\label{fig:figtwo}
\end{figure*}

Here, we aim to measure local material properties at the microscale, subject to 
independently controlled strain amplitude and frequency. We investigate the application 
of steady streaming for microrheology and experimentally demonstrate the measurement 
of kinematic shear viscosity of Newtonian liquids in microfluidic devices. The 
steady streaming regime of an oscillating cylinder \cite{Wang68,Chong,Coenen} is 
dictated by the magnitude of the Reynolds number, $\textrm{Re} = \omega a^2/\nu$, 
and the streaming Reynolds number, $\textrm{Re}_\textrm{s} = \omega s^2/\nu$.
Microscale streaming flows and applications have previously focused on ultrasonic
frequencies and above ($f \geq 10$ kHz), typically induced by the interaction 
between a liquid and surface acoustic waves generated by piezoelectric 
transducers \cite{Friend,Yeo}. The streaming regime associated with high frequencies 
and streaming Reynolds numbers greater than unity ($\textrm{Re}_\textrm{s} > 1$) is 
that of a double streaming layer; an inner driving layer and an outer driven 
layer of the opposite sense. At high frequencies, the inner streaming region is confined 
to a thin layer near the cylinder surface making experimental studies of the inner 
region challenging and therefore less common \cite{Coenen}. By utilizing sub-kHz 
oscillation frequencies, our experiments constrain the streaming Reynolds number to 
always be less than unity ($\textrm{Re}_\textrm{s} < 1$), ensuring that the inner 
streaming layer is comparable in size to the cylinder radius. Because the inner 
boundary layer thickness scales as $\delta \sim \sqrt{\nu / \omega}$, our approach 
characterizes the evolution of the inner streaming boundary layer as a 
function of oscillation frequency with standard particle tracking techniques.


Experiments were performed in microfluidic devices molded in PDMS, consisting of a 
straight channel $20$ mm long, $5$ mm wide, and $200 \ \mu$m tall. Fixed cylindrical 
posts with radii $a$ of $100, \ 200, \ 300,$ and $400 \ \mu$m were manufactured along 
the center of the straight channel. An electroacoustic transducer was used to
externally drive the liquid in a microfluidic device. An oscillatory flow field, 
$U(t)=U_{\infty}\cos(\omega t)$, was setup in the channel over a range of frequencies 
$40 \leq f \leq 1200$ Hz. The oscillation amplitude was independently controlled over 
a range of $s \leq 100 \ \mu$m such that all experiments were in the small-amplitude 
regime $(\epsilon \ll 1)$. The Reynolds number and the streaming Reynolds number 
correspond to a range of $1 \leq \textrm{Re} \leq 1000$ and 
$0.01 \leq \textrm{Re}_\textrm{s} \leq 1$, respectively, for deionized water 
($\nu = 0.949\times10^{-6}$ m$^2$/s) over the entire frequency range investigated here.

Tracer particles, $0.93 \ \mu$m in diameter, were observed at the mid-height of the 
straight channel using bright field microscopy at $10\times$ and $20\times$ 
magnification. Images were acquired using CMOS cameras at sampling frequencies
greater than (high-speed) or at frequencies that are perfect divisors (stroboscopic)
of the oscillatory flow frequency. High-speed imaging provided high fidelity 
observation of the oscillatory flow component while stroboscopic imaging was used 
to characterize the secondary streaming flow component. Experiments were performed at
room temperature, maintained at $20 \ ^{\circ}$C.


\begin{figure}
\centering
	\includegraphics[width=\linewidth]{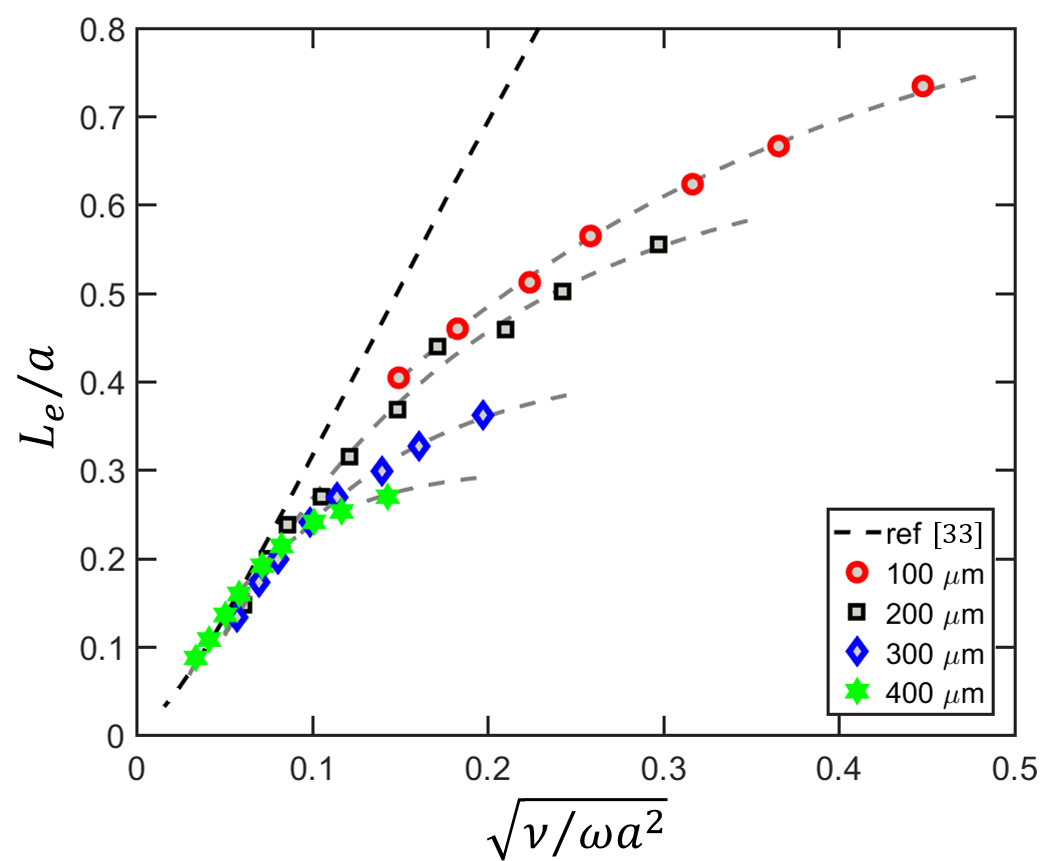}
\caption{Experimental measurements (symbols) of the nondimensional eddy center 
distance from the cylinder surface versus the nondimensional Stokes boundary 
layer for water. Calibration curves (grey dashed lines) for each cylinder radius 
are generated by fitting and exponential function to data points. Experiments
converge to the analytical solution \cite{Holtsmark} (black dashed line) at
high frequencies and large cylinder radii. }
	\label{fig:figthree}
\end{figure}

The oscillatory and steady streaming flow field of an incompressible Newtonian 
liquid near a PDMS cylinder of radius $a$ is illustrated in Figure~\ref{fig:figone}.
The oscillation amplitude is determined from the pathlines of individual tracer 
particles undergoing a single period of oscillation located far from the cylinder, 
approximately $5a$, where the flow is uniform. An example of a minimum projection 
image, captured with high-speed imaging, shows the pathlines of tracer particles 
near the cylinder (Fig.~\ref{fig:figone}, left). When viewed stroboscopically, 
the secondary steady streaming flow component, rather than the oscillatory flow 
component, is observed. The steady streaming flow has a quadrupolar structure, 
consisting of four counter-rotating vortices centered a distance $L_e$ normal 
to the cylinder surface (Fig.~\ref{fig:figone}, right). Flow is directed toward 
the cylinder boundary parallel to the oscillation direction and away from the 
cylinder boundary perpendicular to the oscillation direction (see supplementary 
movie).

The evolution of the steady streaming flow profile with increasing frequency 
for water is shown in Figure~\ref{fig:figtwo}. The four counter rotating vortices 
are identified with tracer particle pathlines (Fig.~\ref{fig:figtwo}, top row). 
The magnitude of the 2D velocity field, obtained using standard particle tracking 
velocimetry (PTV) routines \cite{Ouellette}, clearly shows the location of the 
central vortices with respect to the cylinder boundary (Fig.~\ref{fig:figtwo}, 
bottom row). The eddy center location, $L_e$, decreases in magnitude and approaches 
the cylinder surface as the oscillation frequency is increased.

The dimensionless eddy center location increases monotonically with increasing 
dimensionless Stokes length (Fig.~\ref{fig:figthree}). In this case, the Newtonian 
liquid used was deionized water with $\nu = 0.949\times10^{-6}$ m$^2$/s. For small 
Stokes lengths ($\sqrt{\nu / \omega a^2} \leq 0.1$), the dimensionless eddy center 
increases linearly, in good agreement with the theoretical treatment by Holtsmark 
et al.~\cite{Holtsmark} (Fig.~\ref{fig:figthree}, black dashed line). For larger 
Stokes lengths ($\sqrt{\nu / \omega a^2} \geq 0.1$), the location of the 
dimensionless eddy center diverges from the linear behavior and approaches a 
fixed value of $L_e/a$. The plateauing behavior, for values of 
$\sqrt{\nu/\omega a^2} \geq 0.15$, is attributed to the singular growth in the 
size of the dimensionless inner streaming layer \cite{Bertelsen}. As a 
consequence, the confinement effect of the device boundaries are felt very close
to the cylinder, even for channel width to cylinder radii ratios of greater than 
$20$. The plateau behavior is related to the channel width (in our case is fixed) 
to cylinder radius ratio and therefore largest for the $100 \ \mu$m cylinder and 
smallest for the $400 \ \mu$m cylinder.

Since the behavior was consistent for cylinders of different radii, calibration 
curves were generated by fitting an exponential function,
$L_e/a = A-B \textrm{exp}(\sqrt{\nu/\omega a^2})$, to the experimental data, 
where $A/B$ is approximately unity (Fig.~\ref{fig:figthree}, grey dashed lines). 
Therefore, the kinematic viscosity of a Newtonian liquid can be determined by measuring 
the location of the dimensionless eddy center for given microfluidic device with 
cylinder radius $a$ and angular frequency $\omega$. Microfluidic devices with a 
$200 \ \mu$m cylinder radius provided the largest dynamic range in measurements of 
$L_e$ for the range of operation frequencies. Therefore, the $200 \ \mu$m radius 
cylinder was utilized for the measurements of the kinematic viscosity of Newtonian 
liquids.

\begin{figure}
\centering
	\includegraphics[width=\linewidth]{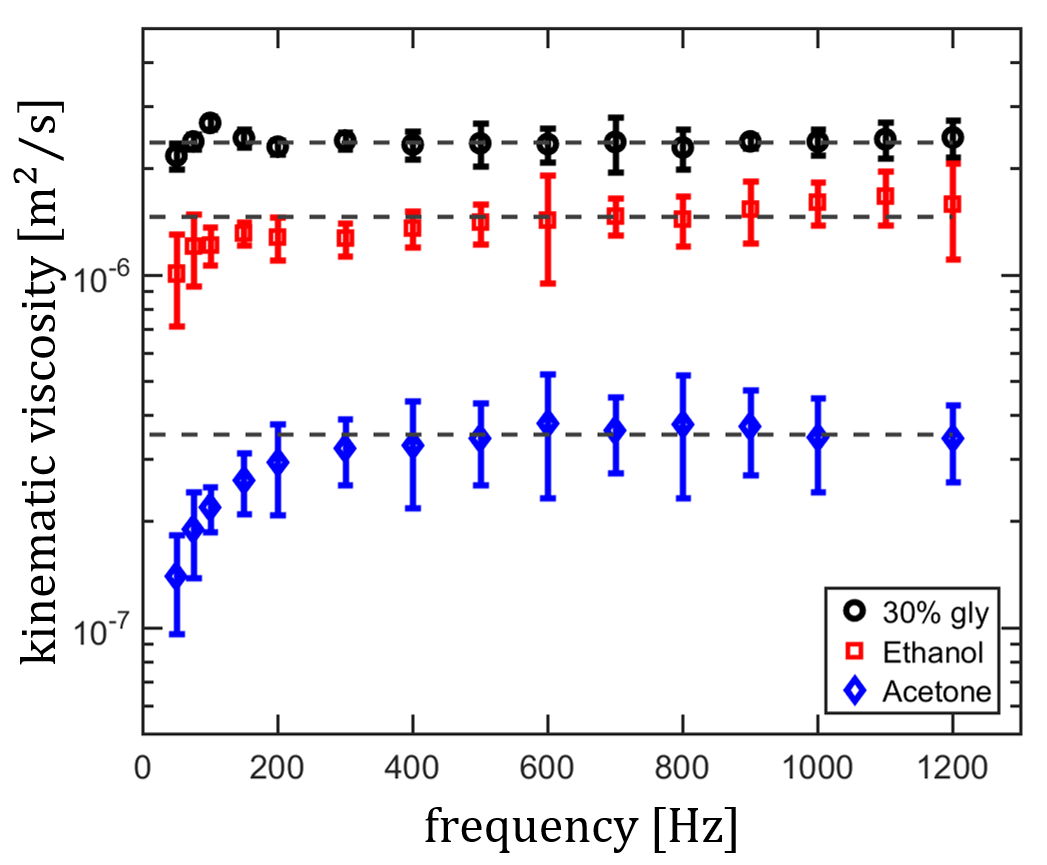}
\caption{Kinematic viscosity measurements of Newtonian liquids using steady 
streaming flows in microfluidic devices with a cylinder radius of $200 \ \mu$m.
Dashed lines represent that average of individual values for frequencies greater
than 300 Hz.}
	\label{fig:figfour}
\end{figure}

The kinematic viscosity of three different Newtonian liquids was determined from 
the steady streaming profiles in microfluidic devices (Fig.~\ref{fig:figfour}). 
The Newtonian liquids used were acetone, ethanol, and an aqueous solution of 
$30\%$ glycerol by weight. For frequencies greater than $200$ Hz, the reported 
measurements of kinematic viscosity are approximately constant, or independent 
of oscillation frequency. Therefore, to obtain a single value of the kinematic 
viscosity, measurements for frequencies of $300$ Hz and greater were averaged 
together. 

The average values of kinematic viscosity are represented by grey dashed 
lines in Figure~\ref{fig:figfour}. The kinematic viscosity of acetone was measured 
to be $3.52\times10^{-7}$ m$^2$/s, which is within $12\%$ of the expected value 
\cite{Howard}. The kinematic viscosity of ethanol was measured to be 
$1.48\times10^{-6}$ m$^2$/s, which is within $2.6\%$ of the expected value \cite{Soliman}. 
Finally, the kinematic viscosity of the $30\%$ glycerol (w/w) solution was measured 
to be $2.37\times10^{-6}$ m$^2$/s, which is within $1.6\%$ of the expected value
\cite{Glycerine}.




The performance of our technique was compared against tabulated values of kinematic
viscosity under similar temperature conditions \cite{Howard,Soliman,Glycerine}. 
The accuracy of our measurements ranged from within $2\%$ to $12\%$, even for 
volatile liquids, such ethanol and acetone. The largest discrepancy, in the case 
of acetone, was primarily due to evaporation and consequent bubble formation at 
the interface with the external driving mechanism. This caused secondary flows 
that interfered and distorted the inner streaming vortices resulting in relatively 
large error, particularly at frequencies less than $200$ Hz. 


In this work, we have demonstrated the use of steady streaming for microfluidic 
viscometry of Newtonian liquids. This optical microrheological technique implements  
independent and precise control of the liquid oscillation amplitude and frequency 
to maintain the streaming Reynolds number to be less than unity 
($\textrm{Re}_\textrm{s} \leq 1$). By utilizing sub-kHz frequencies, we characterize 
the evolution of the inner boundary layer as function of oscillation frequency with 
standard particle tracking and flow visualization techniques of different Newtonian 
liquids. For a given calibrated device, the kinematic viscosity is inferred. 

Like other microrheological techniques, our method has similar advantages such as 
small sample volume, reduced surface effects, and a well-defined channel geometry. 
One additional advantage is a reduced measurement time due to the short transient 
associated with the steady streaming viscometer, which is less than one second. While 
we used high-speed imaging and particle tracking techniques, they are not required 
to characterize the evolution of the inner streaming layer. Measurement of the eddy 
center location, $L_e$, can be determined from visualization of the tracer particle 
pathlines obtained from stroboscopic imaging (Fig.~\ref{fig:figone}, right and 
Fig.~\ref{fig:figtwo}, top row). making this technique accessible to those without 
a high-speed camera. Finally, a major advantage of this technique is the ability to 
quantify volatile and low-viscosity liquids that could be problematic with conventional 
bulk rheoglogical methods. 
This technique could be useful in active microrheology of non-Newtonian fluids
and further extended to investigate rate-dependent material properties and aging 
materials, such as viscoelastic liquids and biological fluids. 


We want to thank Gwynn J. Elfring and Saverio E. Spagnolie for the invitation to 
present this work at the Banff International Research Station workshop, ``Complex 
Fluids in Biological Systems''. We also want to thank Paulo E. Arratia for useful 
discussions and Jonathan B. Freund for feedback on the manuscript. 

\bibliography{streamingrefs}

\end{document}